\documentclass[aps,prl,twocolumn,showpacs,superscriptaddress,groupedaddress,footinbib]{revtex4}
\usepackage[dvips]{graphicx}
\usepackage{subfigure}
\usepackage{epstopdf}
\usepackage{color}
\usepackage{ucs}
\usepackage[utf8x]{inputenc}

\begin{document}
\author{Con Healy$^1$, Sascha Koch$^1$, Carsten Siemers$^2$, Debashis Mukherji$^2$ and Graeme J Ackland}
\affiliation{
 School of Physics and Centre for Science at Extreme Conditions,
University of Edinburgh, Edinburgh EH9 3JZ, UK. e-mail: gjackland@ed.ac.uk \\
$^{2}$ Technische Universitaet Braunschweig, Institut fuer Werkstoffe, 38106 Braunschweig, Germany.
}
\title{Shear melting and high temperature embrittlement: theory and application to machining titanium}
\begin{abstract}

We describe a dynamical phase transition occurring within a shear band
at high temperature and under extremely high shear rates.  With
increasing temperature, dislocation deformation and grain boundary
sliding is supplanted by amorphization in a highly localized nanoscale
band, which allows massive strain and fracture.  The mechanism is
similar to shear melting and leads to liquid metal embrittlement at
high temperature.  From simulation, we find that the necessary
conditions are, lack of dislocation slip systems, low thermal
conduction and temperature near the melting point.  The first two are
exhibited by bcc titanium alloys, and we show that the final one can
be achieved experimentally by adding low-melting point elements:
specifically we use insoluble rare earth metals (REMs).  Under high
shear, the REM becomes mixed with the titanium, lowering the melting
point within the shear band and triggering the shear-melting
transition.  This in turn generates heat which remains localized in
the shear band due to poor heat conduction. The material fractures
along the shear band.  We show how to utilize this transition in the
creation of new titanium-based alloys with improved machinability.

\end{abstract}

\maketitle

There are many atomic-level mechanisms for deformation in metals.
Dislocation motion is the easiest and slowest, while high shear rate
can give rise to twinning, and grain boundary sliding is important in
nanocrystalline materials.  Shear melting is well known in
weakly-bound colloidal systems\cite{shearmeltCol,Belak,weitz} and shearing near
the melting point can induce anisotropic flow\cite{shearmeltTa}.
Liquid metals can cause dramatic embrittlement, but this is generally
believed to be promulgated by existing grain boundaries\cite{liquidmetemb}.  

The shear-melting regime is seldom considered in metallurgy because it
requires temperatures close to melting, even for high shear.  However,
if the shearing {\it also} changes the material composition, then the 
shear band may melt at lower temperature. 

Titanium alloys are notoriously difficult to machine, because the
cutting operation has to be interrupted in order to remove long chips
from the tool\cite{oldTi,micro}.  This problem is typically addressed
by advanced tooling, but there is an alternate approach - designing
alloys which are brittle under manufacturing conditions, but not under
the conditions encountered in use.

In this letter we show that at very high cutting rates the mechanism
of deformation and fracture switches from dislocation motion to a
localized shear-amorphization.  We demonstrate this dynamical phase
transition using an alloy with insoluble intergranular inclusions with
low melting point.  Under high shear, the alloy and inclusion are
mixed to create an alloy with low melting point, which readily
undergoes easy fracture. In cutting tests, we show that the alloy
forms chips with segments separated by localized sheared regions which
readily break off, facilitating machining. The approach is tested on a
series of model alloys, and rare earth metals cerium, lanthanum and
neodymium are shown to be suitable for use as these inclusions.

In machining, three different kinds of chips are known to form:
continuous chips, segmented chips, and completely separated
segments. The chip formation process can be described as follows:
Firstly, the tool penetrates the workpiece and the material is dammed
in front of it. The plastic deformation is then concentrated in
a submicron-scale primary shear zone, leading from the tool tip to the
upper surface of the workpiece. Most of the energy used for the
plastic deformation is transformed into heat in this primary shear
zone. If this heat can dissipate quickly into the areas
surrounding the primary shear zone, the material is deformed
homogeneously leading to the formation of a continuous chip with
constant thickness, and the cutting force remains almost constant. On
the other hand, if the heat cannot dissipate quickly, it is
concentrated in the primary shear zone, the material may locally
soften and the deformation may become even more localized, a feedback effect
further enhancing local heating. In this case, deformation only occurs
in a narrow zone of a few microns (the so-called adiabatic shear band)
which leads to the formation of segmented chips.

The temperature differential between shear band and bulk material
depends strongly on the material properties (e.g. heat conductivity,
yield stress).  Our simple heat-flow analysis (see SM) shows that for
cutting speeds around 10m/s the temperature within a nanoscale shear
band can be raised by $\approx10^3$K before becoming non-adiabatic
(i.e. heat generation in the band matches heat conduction from it),
and that the heating rate is sufficient to reach the melting point in
a sub-micron shear band.

Good machinability requires short, breaking chips,
so for effective cutting, the
deformation must be localized and develop into a fracture. Thus
we first set out to use molecular dynamics to understand how strain can become
localized under high shear rates.

Shear localization implies very high shear rates within the band: the
overall shear rate is increased locally by the degree of localization.
In simulations and in micromachining
experiments\cite{micro,baker} we expect at least 800\% and $10^7 s^{-1}$ strain
and strain rates.  {\it In situ } imaging of this process is impossible,
while continuum modelling of this process is impractical 
since the typical size of the shear band is very much smaller than the
finite elements which can reasonably be used.  Moreover, conventional
constitutive equations
typically assume strain hardening\cite{JohnsonCook}, and can only describe a 
change of deformation mechanism if it is predetermined.

By contrast, atomic-level modelling introduces no such
assumption, and can describe the relevant shear rates, so our
theoretical method of choice for interpreting the experiments is
molecular dynamics.

Previous MD simulations have invesigated nanomachining with nanometer-sized
tools\cite{ye2003molecular,pei2007large}, however we are interested in
a larger lengthscale so are unable to explicitly simulate the tool.
Instead, we carried out molecular dynamics calculations
(MD) on systems which represent the material within the shear
bands, but which are large enough to allow the shear to spontaneously 
localise in one part of the simulation.

The best characterized potentials for bcc metal are fitted to data for
iron, and since we are looking for general features applicable across
a range of difficult-to-machine metals, we choose these for the main
study\cite{mim,fec}.  Similar results were found on repeating the
1500K calculation using a potential fitted to hcp-Ti but
thermodynamically stable as bcc at high temperature\cite{may1999,ti91} 
These are a better representation of
``generic'' metallic bonding than pair potentials such as
Lennard-Jones.

The calculations were carried out using the MOLDY code\cite{moldy}.
The polycrystalline system was set up by first creating a molten
sample of 23 nm$^3$ (1024000 atoms); 48-atom nuclei of bcc crystal
were then introduced at 30 randomly-located points, overlapping liquid
atoms being removed, the sample was then solidified at 1000K, to
produce 30 randomly-oriented grains.  Unlike the rotation
method\cite{nanoswy} our samples contain numerous $\frac{1}{2} < 111
>$ growth dislocations.  This avoids biassing the simulation against
dislocation deformation and work hardening.  The systems were sheared
under strain control at a rate of $2.5\times 10^8s^{-1}$ by rescaling
the y and z coordinates, using the Parrinello-Rahman
lagrangian\cite{pr} which avoids spurious interfaces or shocks, and is
slow enough not to get into the regime of twinning
deformation\cite{shock}. Simulations were carried out at 300K, 600K,
900K, 1200K, and 1500K, which should be interpreted in terms of
proximity to the melting point.

There is a balance between the heat generated by the shearing, and the
rate at which it can be removed by thermal conduction.  Titanium
alloys are notable for their very poor heat conduction (25\% that
of iron), which means that temperature within a shear band may rise
towards the melting point as the energy from the deformation remains
localized.  Our simulations describe the shearing region only, not the
cooler surrounding material, so the temperature is determined by a
Nose thermostat with characteristic time fitted to the experimental Ti
heat conductivity to remove temperature from the system at an
appropriate rate: for an $N-$atom simulation of size $x$, this is
$t=3Nk_B/2Kx$ where $K$ is the thermal conductivity\cite{galloway}.

The target temperature is determined by considering heat flow out of
the shear band, and depends on the material properties.  and shearing
conditions.  A curious result is that the
steady state temperature is broadly independent of the thickness of
the shear bands\cite{shearbandT}. The shear band temperature depends
primarily on the material itself, the machining properties
(e.g. cutting rate) only determining whether there is time to heat the
shear band to this steady state (see Supplemental Materials for details).

\begin{figure}
\fbox{\includegraphics[width=\columnwidth]{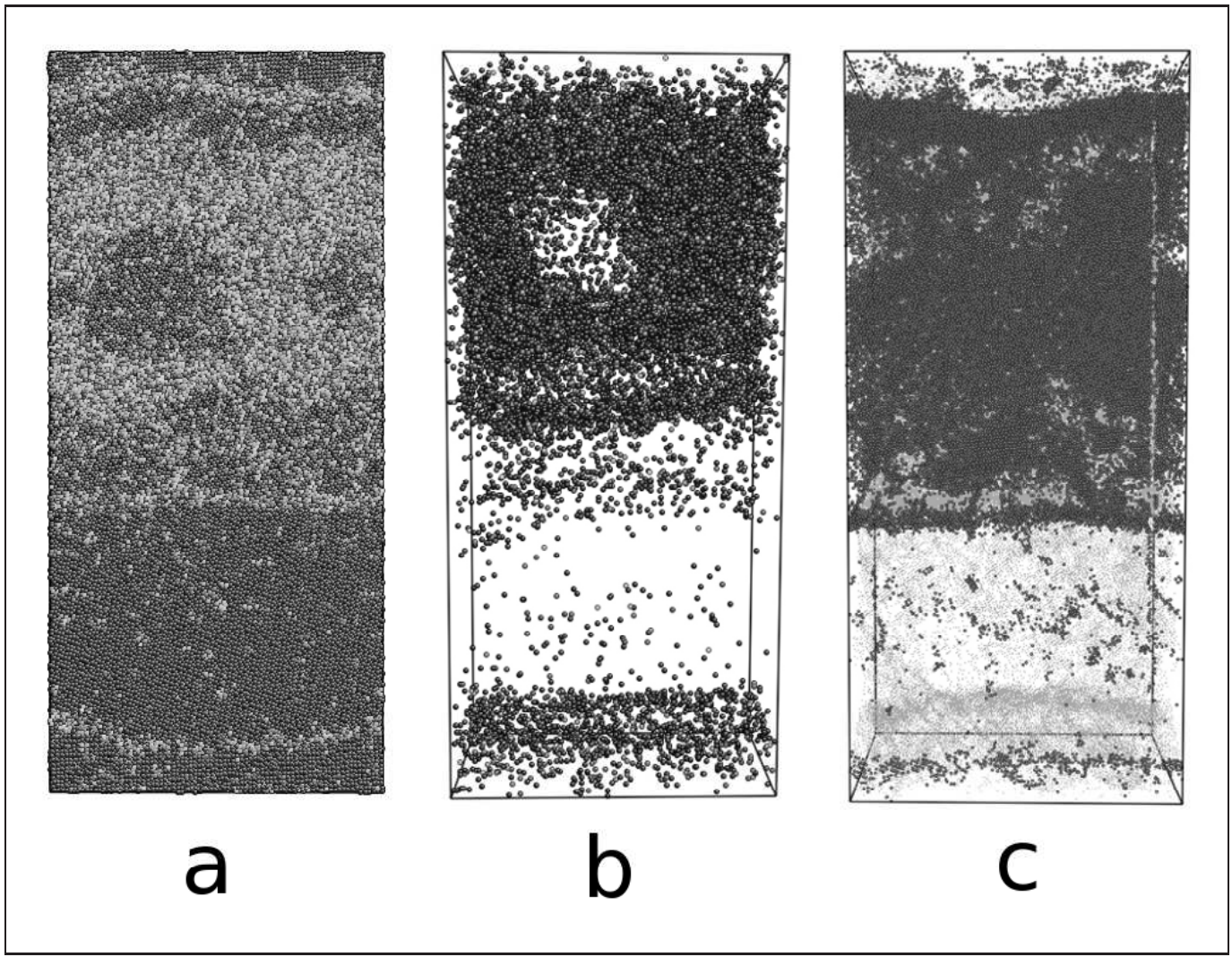}}
\caption{{\bf   Dynamical Phase Transition} Snapshots from molecular
  dynamics simulation at 1500K and 55\% strain at a rate of $2.5\times
  10^8$ showing shear localization and amorphization imaged by Atomeye
  software\cite{atomeye}.  (a) Slice through the system identifies atoms by
  centrosymmetry parameter, blue indicates atoms in close to bcc
  configurations.  (b) 3D projection of only those atoms identified as
  non-crystalline.  The separation between amorphous localized
  shear-band and crystalline region is clear, with the ``hole''
  indicating a nanocrystal ``floating'' in the amorphous zone (c) 3D
  projection showing only those atoms with fewer than 10 of their
  original 14 neighbours, showing that shear and diffusion has been largely 
confined to the amorphous layer.
\label{Kochshear}}
\end{figure}

We have run molecular dynamics simulations of shear at high rates and
various temperatures.  We observe a dynamic transition in the mode of
shear deformation, as illustrated in Fig. \ref{Kochshear} and the
Supplemental movie.  Up to 900K, the deformation proceeds by
dislocation motion which changes the shape of each grain. Some
coarsening of the microstructure can also be seen (Supplemental
Fig.1).
None of these processes generate shear-band
localization.  Above 1400K, the mechanism is similar, until about 50\%
strain, at which point a two phase structure appears spanning the
system.  Even with a few million atoms, all the shear is concentrated
in the localized non-crystalline region.  Within this amorphous
region, there is no ordering by layer as has been seen in smaller,
single-phase simulations\cite{shearmeltTa} small nanocrystallites
survive and move around within disordered surroundings (Supplemental
Fig.2).
 Although there are some boundary-condition
artefacts in the simulation, the central result is that at sufficiently
high temperature, an amorphization mechanism dominates deformation and
causes the shear to becomes localized (Fig. \ref{Kochshear}).  On
cooling such a region the embedded nanocrystallites grow to from a
nanoscale grain structure.  At even higher temperatures the shear band
melts.

Our embedded-atom type potentials\cite{fec,mim} are reliable for
giving a qualitative description of this type of dynamical phase
transition, but cannot give a quantitative description of a real
material. The most reliable way to interpret them is by proximity to
melting point, in which context different MD temperatures represent
different alloy-REM compositions.  The key elements driving
the transition are generic: temperature close to melting point;
inefficiency of the dislocation mechanism to accommodate high shear;
localized adiabatic heating due to poor heat conduction.  These three
conditions are well realized in titanium.

The practical implication of the simulations is that we have predicted
that under cutting conditions this amorphization mechanism causes
shear-localization in a fast-shearing band.  This localization of
energy release maintains a high temperature, which ultimately  
causes melting, associated loss of shear rigidity, and
separation of the segments between shear bands. Alloys with low 
Ti-REM eutectic melting point should show improved machinability.

Alloys with low melting points have other drawbacks, so we attempted
to design a material where the low melting point would manifest itself
only in the shear band during machining (see Supplemental Materials
for sample preparation details). All REMs are almost insoluble in the
Ti matrix at room temperature, and form fine intragranular dispersoids
during crystallization (Supplemental Fig.3). 
Heating the two-phase material melts only the dispersoids, so only the
shear-mixed shear-band region has the reduced melting point.

Other ambient material properties (hardness, moduli etc.) are largely
unchanged whereas the fatigue limit is slightly diminished (by about
10\%) due to the soft, micrometer-size REM-particles (Ref
\cite{CaSi_1} and SM). At the elevated temperatures within the shear
band, we expect that the REMs will become mechanically alloyed with
the matrix, lowering the melting point.  This will trigger the
transition into the regime where amorphization becomes the deformation
mechanism.  We tried various Ti-based matrix materials and note that
our approach was unsuccessful with alloys containing Sn, because
intermetallics like La$_5$Sn$_3$ formed.

\begin{figure}
\includegraphics[width=\columnwidth]{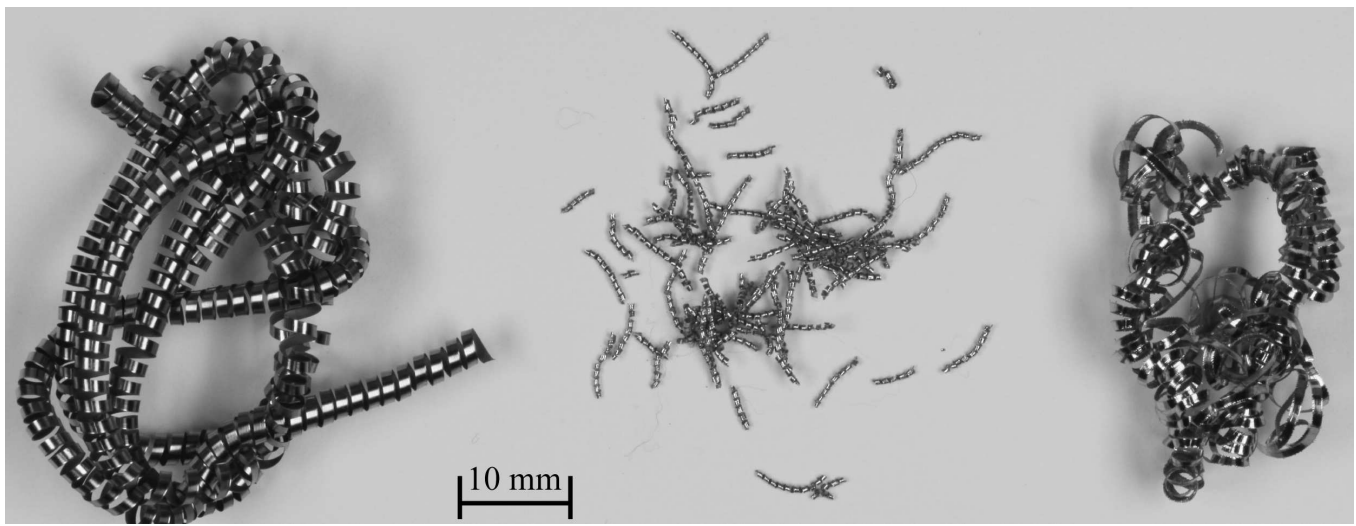}
\caption{{\bf   Chips from conventional turning experiments.} From left to
  right, chips of standard Ti\,6Al\,4V, and with addition of  0.9wt\% La and
  0.9wt\% Er. Chip analyses show that only the La-containing
  material exhibits sufficient shear-band softening
  to fracture the chip. The effect of La on larger scale chip morphology is
  self-evident.
\label{chips}}
\end{figure}

We performed orthogonal high-speed cutting and quick-stop
micromachining experiments \cite{toenshoff,Siemers,procs} which allow for subsequent
analysis of chips and root-chips. In addition, we performed turning
experiments with state-of-the art cutting parameters.  
Using  cutting speeds (v$_c$) between 5 m/s and 100 m/s we find that
alloys with the lower melting-point REMs (Ce, La,
Nd) produce desirable short-breaking chips with completely separated
segments, while the higher melting-point REMs (Y,
Er) deformed conventionally leading to unbroken but segmented chips
(Fig.\ref{chips}). The embrittlement trend follows the eutectic temperature\cite{Massalski}: the shear-band breaks for elements with Ti-REM eutectic
temperature below 1400K, (Ti-Ce: 1070K, Ti-La 1070K, Ti-Nd 1230K), but
compositions with higher eutectic temperatures only give segmented
chips.  (Ti-Er 1600K, Ti-Y 1630K).

\begin{figure}
\includegraphics[width=\columnwidth]{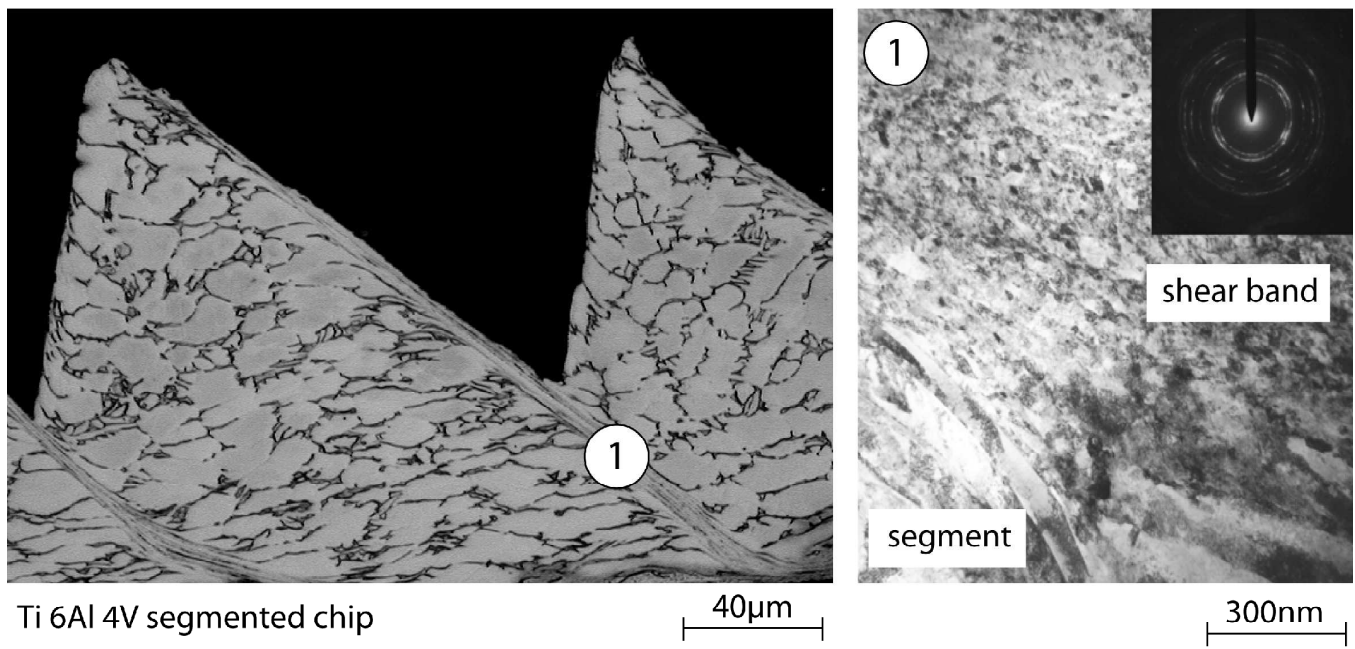}
\caption{{\bf Imaging the localized shear-band.}  Left: Optical microscopy
  image of a Ti\,6Al\,4V segmented chip produced in orthogonal
  high-speed cutting at v$_c$ = 10\,m/s; a$_p$ = 0.1\,mm.  The average
  $\alpha$ grain size in the segments is 20\,$\rm{\mu}$m.  Right:
  Detail from position 1 showing that the shear band consists of
  equiaxed, nano-crystalline $\alpha$-grains. The average grain size
  is about 10\,nm.  The shear band can be clearly distinguished from
  the segment.  The SAD (inset) shows that metastable $\beta$-phase is not
  present.\label{alphachip} Supplemental Fig.6 
  shows the same behaviour in a $\beta$-Ti alloy }
\end{figure}

The post-recovered samples with segmented chips were investigated
using optical microscopy (OM), scanning electron microscopy (FE-SEM)
and transmission electron microscopy (TEM) together with selected area
diffraction (SAD) within the shear band and the relatively undeformed
segment material.  Due to chip fragmentation
only the shear bands of the standard alloy chips could be
analysed by TEM. This reveals that the segmented chips show a
nanocrystalline recrystallized microstructure in the shear band
(Fig. \ref{alphachip}, Supplemental Fig.6).
This has been observed in low-Poission ratio alloys\cite{YangTi}
but here it strongly suggests that the material in the band
has been amorphized and recrystallized, as in the MD calculation. The
grain size outside the shear band is similar to the material prior to
machining, with some additional texture suggesting dislocation
deformation.

Synchrotron X-ray diffraction (See SM) was used to confirm the hcp or
bcc crystal structure\cite{fit2D,CMPR,Karel_1}, depending on alloy type\cite{TegnerPRB}, and the patterns
enable us to detect the dispersoids, intermetallics and to confirm the
absence of internal oxidation. 

Focused X-ray beam patterns obtained from the shear band and
surrounding areas were also collected (Supplemental Fig.5
, only possible in the standard alloys) The
post-machined, segmented samples have both localized shear bands and
relatively undeformed material which we compare
(Fig. \ref{alphachip}).  
The discontinuous nature of the rings
when the beam is focussed on the undeformed region shows that the
grain size is large, similar to the bulk material.  By contrast, the
central pattern with the X-ray beam focussed on the shear band shows
smooth rings show a nanocrystalline structure with grains much smaller
than the beam size.  This is what we expect from a rapid
recrystallization of the amorphous region, and can also be discerned
in the TEM.

The X-ray diffraction also confirms the structure: the former shear
bands show full Bragg rings, corresponding to a nanocrystalline
structure created by deformation melting and dynamic
recrystallization. On the other hand, the patterns taken from the
segments which have not been through the premelting transformation
show Laue spots with pronounced diffuse scattering coming from large
crystallites with preferred orientation \cite{Karel_1}.

In summary, we have found that under high shear rates and temperatures
a transition from dislocation deformation and grain boundary sliding
to shear-band amorphization occurs.  As the temperature increases, the
shear-banded region softens, causing the strain to become more
concentrated within the band.  This leads to a positive feedback as
the stress becomes more localized, depositing still more energy in the
shear band region.  These concepts have been applied to free-machining
titanium alloys using rare earth inclusions (Ce, La, Nd). The high
shear and temperature mixes these insoluble metals, and if the
relevant eutectic alloy has a melting point below the temperature in
the shear band, cohesion is lost and the segments separate.  In
conventional cutting tests, only those REM-doped materials with low
eutectic melting temperatures were found to have short-breaking chips.

This fracture mechanism should be generally applicable to any alloy
producing segmented chips and with low enough thermal conduction that
the shear band can exceed the eutectic temperature for the microsecond
duration of the fracture.  This opens the possibility of creating
free-machining alloys of Ti, Zr, Nb, Ni, Ta, W by addition of
insoluble elements with compartively low melting points. Indeed, we
have also produced a free-machining nickel-base super alloy (Alloy
625) by silver addition \cite{Casi_2}.

\bibliography{Refs}

\noindent {\bf Acknowledgements} This work was supported by a research
grant MAMINA PITN-GA-2008-211536 from the EU-FP7 programme. Beamtime at 
 DORIS, beamline BW5, and PETRA III synchrotron, beamline P07, is gratefully acknowledged.  We would like to thank M. Girelli, A. Gente and B. Zahra for the machining experiments.
The work on Ti\,6Al\,4V alloy has been supported by the Arbeitsgemeinschaft industrieller Forschungsvereinigungen (AiF), Project Number IGF 253 ZN.

 All authors contributed equally to the research. CH, SK and GJA designed and performed the calculations.  CS and DM 
performed the sample preparation, the TEM and XRD experiments.  GJA and CS wrote the paper, with assistance from CH, DM and SK.


\onecolumngrid

\end{document}